\documentstyle[11pt,paspconf,epsf,rotating,twoside,psfig]{article}
\def\beq{\begin{equation}}
\def\eeq{\end{equation}}
\def\l{\label}
\def\bea{\begin{eqnarray}}
\def\eea{\end{eqnarray}}

\renewcommand{\thefigure}{\arabic{figure}}
\renewcommand{\thetable}{\arabic{table}}
\def\ltsima{$\; \buildrel < \over \sim \;$}
\def\simlt{\lower.5ex\hbox{\ltsima}}
\def\gtsima{$\; \buildrel > \over \sim \;$}
\def\simgt{\lower.5ex\hbox{\gtsima}}

\begin{document}
\title{  Detection of non-random patterns in 
large-scale structure}  

\author{ Riccardo Valdarnini}
\affil{ SISSA Via Beirut 2-4 34014 TRIESTE}

\begin{abstract}
A new method for analyzing the morphological features of point patterns 
 is presented. The method
is taken from the study of molecular liquids, where it has been introduced for
making a statistical description of anisotropic distributions. The statistical 
approach is based on the spherical harmonic expansion of angular correlations.
\end{abstract}

\section{ THE METHOD}
\label{sec:meth}
 
 In Valdarnini (2001) I propose an alternative method for analyzing the 
clustering morphology which is a generalization of the 2-point function  
$\xi(r)$ (Peebles \& Hauser 1974) and is based on the spherical harmonic 
analysis.
The method is drawn from molecular dynamics simulations,  where it 
has been introduced 
for studying orientational order of supercooled liquids and
metallic glasses (Wang \& Stroud 1991).
Let us consider a system of $N_p$ particles.
  The $i-th$ particle has coordinates $\vec r_i$, in 
  an arbitrary reference frame.
For a specified cutoff radius $R_c$ all the particles such that 
 $|\vec r_i-\vec r_j|< R_c$ are neighbors of $i$.
The line joining $i$ to one of the $j$ is termed a {\it bond}.
The angular coordinates of the vector $\vec \Delta_{ji}\equiv \vec r_j - \vec r_i $ are $\theta_j, \phi_j$ and the quantity 
\smallskip
\beq
Q_{lm}(\vec r_i)=\sum_{j \ne i} Y_{lm}(\theta_j,\phi_j),
\l{eq:defQ}
\eeq
is the coefficient of the spherical harmonic expansion of the angular density
of the bonds associated with the particle $i$.
In Eq. \ref{eq:defQ}, and hereafter, summation is understood over all
 particles $j$ of the distribution such that 
 $|\vec r_i-\vec r_j|< R_c$.

The coefficients $Q_{lm}(\vec r_i)$ are defined as the 
bond-orientational order parameters and they
can be drastically changed by a rotation of the reference systems.
A natural quantity to consider , which is rotation invariant, is
\beq
Q_l(\vec r_i)=\sqrt {{{4\pi}\over{2l+1}} \sum_{m=-l}^{m=l}
Q^{\star}_{lm}(\vec r_i) Q_{lm}(\vec r_i)}.
\l{eq:Qla}
\eeq
Using the addition theorem for the spherical harmonics, the expression 
simplifies to
\beq
Q_l(\vec r_i)=\sqrt { \sum_j \sum_k P_l(\gamma_{jk})},
\l{eq:Qlb}
\eeq
where $P_l$ is the Legendre polynomial, 
$$
\gamma_{jk}\equiv cos(\theta_{jk})=\vec \Delta_{ji}\cdot\vec\Delta_{ki}/
(|\vec \Delta_{ji}||\vec \Delta_{ki}|)\nonumber
$$
is the angle between two bonds and  the summations in  Eq.
\ref{eq:Qlb} are then independent of the chosen frame.
An  useful quantity is the auto-correlation 
   $ G_l(r)$  of the coefficients
$Q_l(\vec r_i)$. The function $G_l$ is defined as follows:
for all of the $M_p$ pairs $(i,k)$, such that 
$|\vec r_i-\vec r_k|=r \pm \Delta r$, where $\Delta r$ is the 
thickness of the radial bin, then
  $G_l(r) $ is the sum over all of these pairs 

\beq
G_l(r)={{1}\over{M_p}} \sum_i \sum_k{{4\pi}\over{2l+1}} 
\sum_{m=-l}^{m=l} Q^{\star}_{lm}(\vec r_i) Q_{lm}(\vec r_i+\vec r).
\l{eq:GL1}
\eeq

This equation can be greatly simplified: let
 ${j}$ be the set of neighbors of the particle $i$ 
and  ${p}$ that of the particle $k$, which satisfy
 $|\vec r_j-\vec r_i|< R_c$  and $|\vec r_p-\vec r_k|< R_c$. 
Then the summation becomes
\beq
G_l(r)={{1}\over{M_p}} \sum_i \sum_k \sum_j \sum_p P_l(\Gamma_{jp}),
\l{eq:GL4}
\eeq
with {\bf $\Gamma_{jp}$} being the angle between $\vec \Delta_{ji}$ and 
$\vec \Delta_{pk}$.  The summation over the pairs  is   
 $\sum_i \sum_k$, with the sum over the particles $k$ 
only for those particles with $|\vec r_i-\vec r_k|=r \pm \Delta r$.  

The effectiveness of the statistical analysis in quantifying
clustering morphology is studied (Valdarnini 2001) by applying the statistical 
estimator $G_l$ to point distributions produced by an ensemble of 
cosmological $N-$ body simulations with a CDM spectrum.
The results shown that the statistical method
defined by the function $G_l(r)$ can be used to analyze the
clustering morphology produced by gravitational clustering in a 
quantitative way.
The function $G_l(r)$ describes
anisotropies in the clustering distribution by measuring the degree of 
correlation between the angular densities as seen from two different 
observers separated by $r$.
$G_l(r)$ can then be considered a statistical measure of clustering
patterns, with different scales probed by varying the input
parameters $l$ and $R_c$. 
With large redshift surveys becoming available in the next few years, the
proposed statistical method appears as a promising tool for analyzing
patterns in the galaxy distribution.

\end{document}